\documentclass[aps,prl,showkeys,twocolumn,superscriptaddress,floatfix]{revtex4-2}
\usepackage[utf8]{inputenc}
\usepackage{verbatim}
\usepackage{amsmath}
\usepackage{mathtools}
\usepackage{gensymb}
\usepackage{amssymb}
\usepackage[pdftex]{graphicx}
\newcommand{\figurescale}{1}
\usepackage{soul}
\usepackage{xcolor}
\usepackage[normalem]{ulem}

\usepackage{soul}

\begin{document}

\title{The morphology and interface structure of titanium on graphene}

\affiliation{Division of Physical Sciences, College of Letters and Science,
University of California, Los Angeles, California 90095, USA}
\affiliation{Department of Materials Science and Engineering, Massachusetts Institute of Technology, Cambridge, Massachusetts 02139, USA}
\affiliation{Department of Mechanical Engineering and Materials Science, University of Pittsburgh, Pittsburgh, PA 15261, USA}
\affiliation{Electrical and Computer Engineering Department,
University of California, Los Angeles, California, 90095, USA}

\author{Joachim Dahl Thomsen}
\email{j.thomsen@fz-juelich.de}
\affiliation{Division of Physical Sciences, College of Letters and Science, University of California, Los Angeles, California 90095, USA}
\affiliation{Department of Materials Science and Engineering, Massachusetts Institute of Technology, Cambridge, Massachusetts 02139, USA}

\author{Wissam A. Saidi}
\affiliation{Department of Mechanical Engineering and Materials Science, 
University of Pittsburgh, Pittsburgh, PA 15261, USA}

\author{Kate Reidy}
\affiliation{Department of Materials Science and Engineering, Massachusetts Institute of Technology, Cambridge, Massachusetts 02139, USA}

\author{Jatin J. Patil}
\affiliation{Department of Materials Science and Engineering, Massachusetts Institute of Technology, Cambridge, Massachusetts 02139, USA}

\author{Serin Lee}
\affiliation{Department of Materials Science and Engineering, Massachusetts Institute of Technology, Cambridge, Massachusetts 02139, USA}

\author{Frances M.~Ross}\email{fmross@mit.edu}
\affiliation{Department of Materials Science and Engineering, Massachusetts Institute of Technology, Cambridge, Massachusetts 02139, USA}

\author{Prineha Narang}\email{prineha@ucla.edu}
\affiliation{Division of Physical Sciences, College of Letters and Science, University of California, Los Angeles, California 90095, USA}

\date{\today}

\begin{abstract}
\textbf{\abstractname:} Titanium (Ti) is an adhesion and contact metal commonly used in nanoelectronics and two-dimensional (2D) materials research. However, we find that dramatically different film morphology can result when Ti is deposited on graphene (Gr), depending on the experimental conditions. Through a combination of transmission electron microscopy, atomic force micoscopy, and Raman spectroscopy, we show that the most critical parameters are the number of Gr layers, the nature of the Gr support, and the deposition temperature. Monolayer Gr is particularly distinctive in both its island morphology and its high defect density in Gr, compared to bilayer or thicker Gr. We attribute these results to the structural and mechanical differences between monolayer and thicker Gr flakes, where monolayer Gr is more flexible, exhibits larger surface roughness and therefore lower Ti diffusivity, and is more easily damaged. 
This is supported by \textit{ab initio} density functional theory calculations which suggest that differences in the Ti interaction between monolayer and thicker Gr are due to extrinsic factors such as surface roughness.
Our results highlight the extreme sensitivity of Ti morphology on Gr to processing and substrate conditions, allowing us to propose design rules for controlling Ti-Gr interface properties and morphology and to discuss the implications for other technologically relevant metal deposition processes.
\end{abstract}


\maketitle

Control over the properties of metal thin films is essential for the functionality of electronic, optical, and magnetic devices. As device miniaturization progresses, the need for atomic-level precision in metal thin film structure becomes increasingly important \cite{zhang1997atomistic}. Factors such as the nucleation density and growth morphology influence thin-film stress and grain boundary density, and hence the electrical properties of both the metal film and the interface 
\cite{michely2012islands, ratsch2003nucleation, doerner1988stresses}.

\begin{figure*}[]
	\scalebox{\figurescale}{\includegraphics[width=0.67\linewidth]{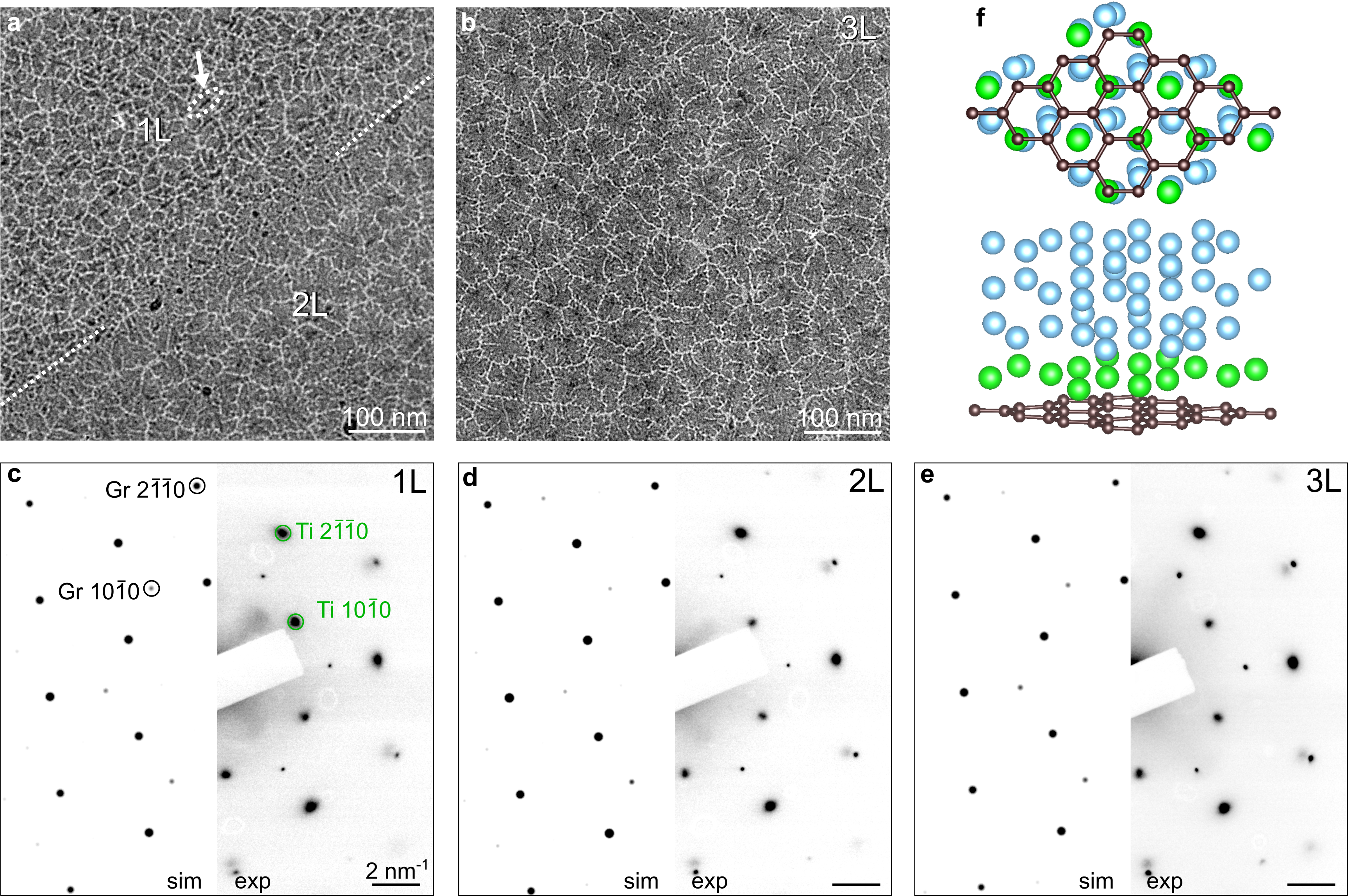}}
	\caption{\label{RT}
		\textbf{The effect of Gr layer number on the Ti-Gr interface.} 
		\textbf{(a, b)} TEM images of suspended (a) 1L and 2L Gr, (b) 3L~Gr. Both images are from the same sample with 1~nm Ti deposited. A step edge separates the 1L and 2L regions in (a), indicated with a white dotted line. The white arrow indicates a dark elongated Ti island which is further described in Fig.~S2. 
		\textbf{(c-e)} The right halves show experimental diffraction patterns from  suspended 1L, 2L, and 3L Gr, respectively. The left halves show simulated diffraction patterns from the predicted structures obtained from DFT (see also (f)). Miller-Bravais indices are indicated in (c). All scale bars are 2~nm$^{-1}$.
		\textbf{(f)} DFT calculations of the atomic interface structure between 1L Gr and the Ti(0001) surface. To highlight the interface structure, the Ti atoms at the interface are shown in green and the rest of the Ti atoms are shown in blue, with carbon atoms shown in brown. 
		}
\end{figure*}

On two-dimensional (2D) materials, the metal thin film quality is particularly critical for applications where the properties and performance of the system depend on the 2D material/metal interface. Notable examples include contact resistance to 2D materials \cite{shen2021ultralow, wang2019van}, proximity-induced superconductivity between topological insulators and superconductors \cite{flototto2018superconducting}, and spintronics applications \cite{roche2015graphene}. On van der Waals (vdW) surfaces, the formation of a metal thin films follow steps of initial nucleation, growth of islands, and coalescence \cite{ruffino2017review}. Factors that affect the diffusion distance or nucleation sites may thus impact the properties of the thin film that eventually forms. Various factors determine the metal-2D material interaction, including surface roughness that influences diffusion properties \cite{thomsen2022suspended,liu2013slow}, reactions of the metal with the substrate \cite{schauble2020uncovering} or background gases during deposition \cite{mcdonnell2016mos2}, contamination from the ambient such as water vapor or hydrocarbons affecting metal morphology and diffusion properties \cite{metois1978experimental}, and contamination from preparation of the original surface \cite{lin2012graphene,zan2011metal}. Understanding how the extrinsic factors influence metal diffusion, nucleation, and growth is necessary for fabricating well-controlled films with good structural and electronic properties. 

In graphene (Gr), a material with ultrahigh charge carrier mobility \cite{wang2013one,banszerus2015ultrahigh}, controlling thin film properties and metal-2D interactions is especially critical in applications such as electronics \cite{schwierz2013graphene, cusati2017electrical}, spintronic devices 
\cite{han2014graphene}, wide-band photodetection 
\cite{goossens2017broadband}, ballistic devices 
\cite{auton2016graphene}, and hydrogen storage 
\cite{mashoff2013hydrogen}. Applications involving other 2D materials may depend equally critically on the quality of metal-2D contacts \cite{du2014rm}. 
In spite of its importance, forming consistent metallic contacts to Gr remains challenging \cite{cusati2017electrical}. Understanding the intrinsic interaction between metals and Gr is essential for improving the processes that form high-quality and homogeneous metallic films while minimizing defect formation.

Of particular interest is titanium (Ti), which is widely used as an adhesion layer in nanoelectronics due to the strong chemical interactions it can have with substrates, and which shows minimal interdiffusion into noble metals such as gold \cite{todeschini2017influence, schulman2018contact}. Furthermore, Ti islands are important for optoelectronic applications, such as the bottom-up fabrication of plasmonic nanoparticle arrays for optical devices \cite{wang2018rich} and catalysis \cite{mascaretti2022challenges}. However, the exact nature of the interaction between Gr and Ti remains debated. For instance, literature is divided on whether defects form in Gr after Ti deposition \cite{matruglio2016contamination, hsu2014surface, xu2018properties}. In addition, it is possible that Ti and Gr react during deposition, forming TiC \cite{gong2014realistic, politou2015transition}. However, it has also been suggested that TiC formation is due to a reaction between Ti and background gases or contamination in the vacuum system and not a reaction between the Ti and Gr \cite{freedy2018unraveling}. 
These varying reports suggest that the Ti-Gr interface properties are highly sensitive to experimental conditions, requiring systematic measurements to explore the impact of contributing factors.

To elucidate the fundamental factors determining diffusion, nucleation, and growth of Ti on Gr, and to distinguish between underlying effects such as substrate, strain, and roughness, it is useful to consider metal deposition on suspended Gr, i.e., Gr not in direct contact with a substrate. This approach enables the study of intrinsic metal-Gr interactions without extraneous effects from the substrate. Suspended Gr is itself promising for applications in areas such as filtration \cite{surwade2015water}, DNA sequencing \cite{heerema2016graphene}, and catalyst support \cite{antolini2012graphene, lv2011open, julkapli2015graphene}. On suspended Gr, investigations of the early-stage growth of Ti reveal that growth takes place in the Volmer-Weber island growth mode and the initial island layers are strained to accommodate the lattice mismatch between Ti and Gr \cite{zagler2023interface}.

Here, we examine the interaction between Ti and Gr systematically, as a function of Gr layer thickness, substrate, and temperature. This allows us to clarify the influence of these different factors on the Ti-Gr interaction and the resulting Ti morphology and Ti-Gr interface structure. 
We find that monolayer Gr behaves dramatically different compared to thicker Gr in two key respects: the Ti morphology and a large defect density that is present in the Gr after Ti deposition. We identify surface roughness--which in turn is affected by the choice of substrate and Gr thickness--as an important factor controlling the Ti morphology and defect formation in Gr.
We present these results as follows: (1) We describe the dependence of Ti island morphology and nucleation characteristics on Gr thickness, through a combination of TEM and \textit{ab initio} density functional theory (DFT) calculations; (2) We determine the effect of deposition temperature on Ti morphology; (3) We consider substrate effects by comparing results for suspended Gr and SiN-supported Gr; (4) We investigate defect formation during Ti deposition by Raman spectroscopy; (5) We summarize the effects of Gr thickness, temperature and substrate on Ti morphology and defect formation and discuss strategies aimed at forming Ti-Gr interfaces with minimized defect density and large expitaxial Ti grains.

\begin{figure}[h!]
	\scalebox{\figurescale}{\includegraphics[width=1\linewidth]{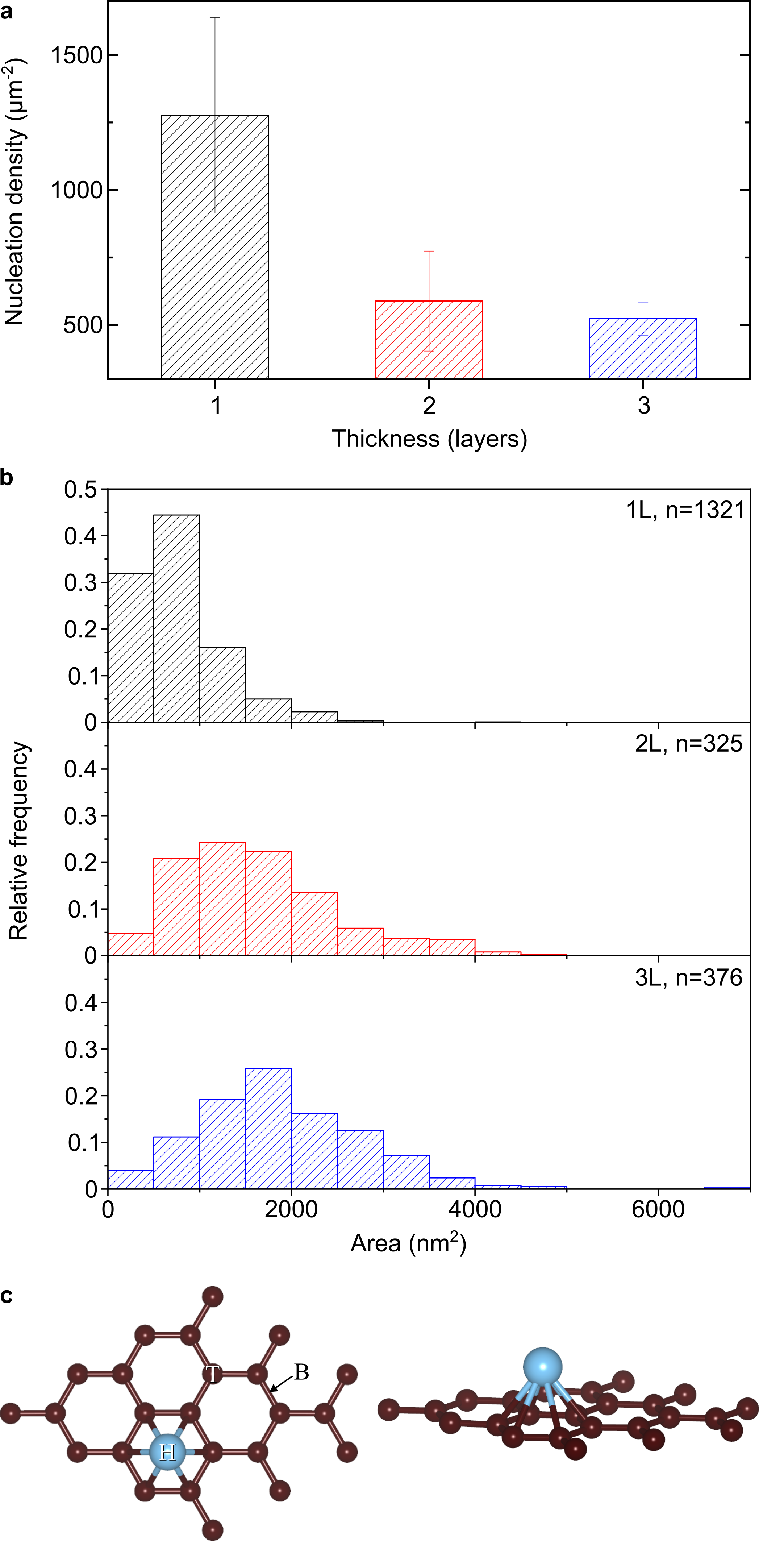}}
	\caption{\label{fig:nucdens}
		\textbf{Ti nucleation characteristics on suspended Gr.} 
		\textbf{(a)} Nucleation density as measured from the images shown in Fig.~S3. The total areas measured for 1L, 2L, and 3L Gr were 0.90, 0.55, 0.73~µm$^2$, respectively. 
		\textbf{(b)} Histograms of Ti island areas for 1L, 2L, and 3L Gr. \textit{n} is the number of measured islands.
        \textbf{(c)} Top and side view, respectively, of the most stable adsorption site for Ti on Gr, the hollow (H) site. Also indicated are the top (T) and bridge (B) sites.} 
\end{figure}

\section{Results/Discussion}
\subsection{Graphene thickness dependence on Ti-island morphology and nucleation characteristics}
Figure~1(a, b) shows the results of depositing, by evaporation, a 1~nm thick Ti layer at room temperature onto 1L, 2L and 3L suspended Gr. Optical microscopy images of the Gr before and after transfer to the TEM sample carrier, shown in Fig.~S1, allow us to correlate the location of images recorded in TEM with the thickness of the Gr at that point. We find that the Ti forms irregularly shaped islands on Gr, but in spite of their irregular appearance these islands are aligned epitaxially. An irregular island shape is expected given the low homologous temperature during these depositions \cite{michely2012islands}, {$\sim$}0.15 $T_{\mathrm{M}}$, where $T_{\mathrm{M}}$=1941~K is the melting temperature of Ti. The darker, elongated Ti islands, such as that indicated by the white arrow in Fig.~1(a), appear to be locally thicker Ti (Fig.~S2). Imaging at high magnification does not reveal visible damage to 1L Gr (Fig.~S2); below we show different results when the deposition is performed at high temperature.

The uniformity of the Ti film suggests that the growth is not dominated by a low density of heterogeneous nucleation sites such as surface contaminants, a conclusion supported by the temperature dependence of nucleation density discussed below. Investigating the intrinsic interaction between metals and 2D materials generally requires that the 2D material is free from adsorbed contaminants, particularly water and polymer residues from the transfer process. The samples shown in Fig.~1 were prepared using a transfer process combined with sample annealing and metal deposition, the latter two conducted in ultra-high vacuum (UHV) conditions, to create the suspended and supported Gr samples on which we deposit Ti. This process results in the near-complete elimination of contamination on suspended Gr \cite{thomsen2022suspended, reidy2022mechanisms}. Even so, it is important to minimize the time between Gr annealing and Ti deposition, because the Gr surface can re-contaminate even in UHV conditions \cite{thomsen2022suspended, reidy2023perspectives}. Further details of sample preparation and Ti deposition are described in \textit{Methods}.

We now discuss the epitaxial arrangement displayed by the Ti on Gr. The Ti $\langle2\bar{1}\bar{1}0\rangle$ directions are parallel to the Gr $\langle10\bar{1}0\rangle$ directions as shown in the diffraction patterns in Fig.~1(c-e), and the Gr~[1000] surface normal is parallel to the Ti~[1000] surface normal. 

Given the size of the selected area aperture used to acquire the diffraction patterns (diameter 520~nm), each pattern includes signals from tens or hundreds of Ti islands. These patterns show that each separate Ti island grows in identical alignment with the Gr. This crystal alignment between Ti and Gr is consistent with predictions from \textit{ab initio} calculations that we performed (see \textit{Methods}), as well as with results from literature \cite{hsu2014surface, zagler2023interface, reidy2022mechanisms, fonseca2017graphene}. Figure~1(f) shows the expected atomic interface structure between Ti and Gr. 

The nucleation density of Ti islands determines parameters such as density of grain boundaries and film stress and has a direct impact on the properties of the resulting thin films \cite{zhang1997atomistic, michely2012islands, ratsch2003nucleation, doerner1988stresses}. For the Ti islands on suspended Gr in Fig.~1, we find that changes in nucleation density and island size are visible depending on the Gr layer number. Each area shown in Fig.~1(a,\,b) received the same Ti flux, yet the island size increases and nucleation density decreases with increasing Gr layer number. Figure~2(a) shows this quantitatively in a plot of the Ti island nucleation density for 1L-3L Gr following the deposition of 1~nm Ti at room temperature, while Fig.~2(b) shows histograms of the island areas. From this data we conclude that the difference in nucleation density between 1L Gr and 2L/3L Gr is statistically significant. Figure~S3 shows the details of the images from which we derived this data and the method used to determine the individual island boundaries. Details of these measurements are given in \textit{Methods}.

\begin{table*}[t]
\centering
\begin{tabular}{p{2.5cm} p{3cm} p{4cm} p{3.5cm} p{3cm} p{0cm}} \hline
\centering \textbf{Gr thickness (L)} &  \centering \textbf{Total adsorption energy (eV)} &  \centering \textbf{Short-range adsorption energy (eV)} & \centering \textbf{vdW adsorption energy (eV)} & \centering\textbf{Interface energy (eV/Å$^2$)} & \\ \hline 
\centering 1 & \centering -2.21 & \centering -1.89 & \centering -0.32 & \centering -0.148 &\\  
\centering 2 & \centering -2.33 & \centering -1.88 & \centering -0.45 & \centering -0.161 &\\
\centering 3 & \centering -2.36 & \centering -1.91 & \centering -0.45 & \centering -0.163 &\\ \hline
\end{tabular}
\caption{Results from DFT calculations. The total, short-range, and vdW adsorption energies, listed in units of eV, are for a single Ti atom on the hollow site. The interface energy (eV/Å$^2$) is calculated for a Ti and Gr slab as outlined in Methods.}
\label{tab:1}
\end{table*}

We use this data in a mean-field diffusion theory to estimate the diffusion constant of Ti on suspended Gr. Assuming a constant flux of Ti from the evaporator to all regions of the sample, the nucleation density, $N$, can be expressed as \cite{venables1983surface}

\begin{equation}
    N \propto \left( \frac{1}{D} \right)^{\frac{i}{i+1}}
\end{equation}

where $D$ is the diffusion constant and $i$ is the critical nucleus size. Equation 1 assumes that individual islands have not yet coalesced. Our islands are in close proximity but we can still distinguish between individual islands and determine from their shape and contrast that each island is a result of a single nucleation event. We use values $N_{\mathrm{1L}}$ = 1280~µm$^{-2}$, $N_{\mathrm{2L}}$ = 590~µm$^{-2}$, and $N_{\mathrm{3L}}$ = 520~µm$^{-2}$ (Fig.~2). The critical nucleus size of metals can generally be considered small \cite{venables1983surface, brune1998microscopic}, so we assume that $i$=1 or 2. These values lead to results $D_{1}/D_{2}$  = 0.07 and 0.18, respectively, and $D_{1}/D_{3}$  = 0.04 and 0.13, for $i$=1 and 2, respectively. Thus, the measurement shows that the diffusion coefficient of Ti on suspended Gr is 6-14x larger for 2L Gr compared to 1L Gr, and 8-25x larger for 3L Gr compared to 1L Gr.   

Such layer-dependent metal diffusion properties have previously been observed for Au on Gr \cite{thomsen2022suspended, zhou2010thickness, luo2010size, zhou2011high, zhou2013transformation}. The Gr layer dependence for Au diffusion was attributed to variations in surface roughness with layer number \cite{thomsen2022suspended, liu2013slow}. The underlying mechanism involves the diffusion coefficient being reduced or increased by compressive or tensile strain, respectively \cite{ratsch1997strain, brune1995effect}. As a result, regions with compressive strain act as effective nucleation sites due to lower metal diffusivity in those areas. Suspended 1L Gr exhibits a root mean square roughness of about 0.1~nm \cite{thomsen2017suppression, gao2014thermomechanics, ramirez2016anharmonic}, which decreases as the layer thickness increases \cite{meyer2007roughness}. Extending these ideas from Au to Ti, the thickness dependence on Ti island size in suspended Gr can be understood consistently in terms of thickness-dependent surface roughness. 

To aid in understanding the differences $N$ and interaction as a function of Gr layer number, we performed \textit{ab initio} DFT calculations. As we show below, these calculations predict a slightly increasing adsorption energy for single Ti atoms on Gr, and an increasing interface energy for Ti slabs on Gr, with increasing Gr thickness. Since experimental results in Fig.~1 and 2 demonstrate the opposite trend, that the Ti-Gr interaction decreases with increasing Gr thickness, this comparison of experiment with DFT results suggests that extrinsic factors such as surface roughness that are not considered in the DFT calculations dominate the layer-dependent Ti-Gr interaction. 

Our calculations first consider the single atom adhesion energy of Ti, which was calculated using $E_{\mathrm{ads}}=E_{\mathrm{(Gr/Ti)}}-E_{\mathrm{Gr}}-E_{\mathrm{Ti}}$, where $E_{\mathrm{(Gr/Ti)}}$, $E_{\mathrm{Gr}}$, and $E_{\mathrm{Ti}}$ are the energies of the 1L-3L Gr with adsorbed Ti, pristine Gr, and isolated Ti atoms with a spin configuration consistent with Hund’s rules, respectively.  Given the symmetry of the Gr lattice, a Ti adatom can adopt three different adsorption configurations on the top, bridge, and hollow sites, as shown in Fig.~2(c). The calculations show that, irrespective of the number of Gr layers, the hollow site is the most stable adsorption site for Ti with $E_{\mathrm{ads}}$ for 1L-3L given in Table~1. Adsorption energies for the other sites are given in Table~S1. The values are consistent with previous calculations on 1L Gr \cite{valencia2010trends} and are also consistent with adsorption trends of metals on other 2D surfaces such as MoS$_2$ \cite{saidi2015trends, saidi2015density}.

Table 1 shows that the Ti adsorption energy decreases (i.e.\ stronger interaction) by $\sim$0.1~eV/Ti~atom as the Gr thickness increases from 1L to 2L, and then decreases further by $\sim$0.03~eV/Ti~atom from 2L to 3L. Examining the contributions of the short-range and long-range (vdW) adsorption energy shows that the slight increase in the total adsorption strength as Gr thickness increases originates from increased vdW interactions. The short-range interactions, as captured by the energy without the vdW corrections, show the opposite trend.

We next consider the interface energy between Gr and Ti. The interface energy was found by minimizing the misfit strain between Ti (0001) and Gr, with structures as shown in Fig.~1(f). Simulated diffraction patterns from such structures match experimental diffraction patterns well (Fig.~1(c-e)). The results for the interface energies are given in Table 1. As is the case for single adsorbed Ti atoms, we find that the interface energy decreases slightly with Gr thickness due to vdW interactions, and hence predicts a stronger Ti-Gr interaction with increasing Gr thickness. 

Our overall conclusion from the \textit{ab initio} calculations for both single Ti atoms and slabs of Ti is that the intrinsic interactions between Gr and Ti are not very sensitive to the Gr thickness, and that a slightly stronger interaction is expected when the Gr thickness increases. Our experimental observation that $N$ decreases with increasing layer thickness (Fig.~1) implies a larger Ti diffusivity on thicker Gr. This might suggest a weaker Ti-Gr interaction with increasing thickness, but this was not reproduced by DFT calculations. We therefore conclude that the experimental differences in $N$ for 1L-3L suspended Gr are due to an extrinsic factor, with surface roughness being a plausible origin. 

In summary, we find that increasing Gr thickness leads to higher Ti diffusivity on the Gr surface, resulting in epitaxial islands with a lower nucleation density and hence larger size. We suggest, based on the comparison with DFT calculations, that the observed Gr thickness dependence is due to surface roughness which decreases with increasing Gr thickness, analogous to previous observations for Au on Gr \cite{thomsen2022suspended}. We summarize our findings in Table 2 along with other data that will be discussed next.

\subsection{Effect of deposition temperature on Ti morphology}
As a means of Ti morphology control, we now discuss samples where Ti has been deposited on Gr at elevated temperatures. Figure 3(a) shows a TEM image of 1L and 2L Gr after 1 nm Ti deposition while heating to 400\,$\degree$C. Additional images of 1L, 2L, and 3L Gr with Ti deposited at room temperature, 300\,$\degree$C, and 400\,$\degree$C, corresponding to 0.30-0.35 $T_{\mathrm{M}}$, are provided in Fig.~S4-S8.

\begin{figure*}[t]
	\scalebox{\figurescale}{\includegraphics[width=1\linewidth]{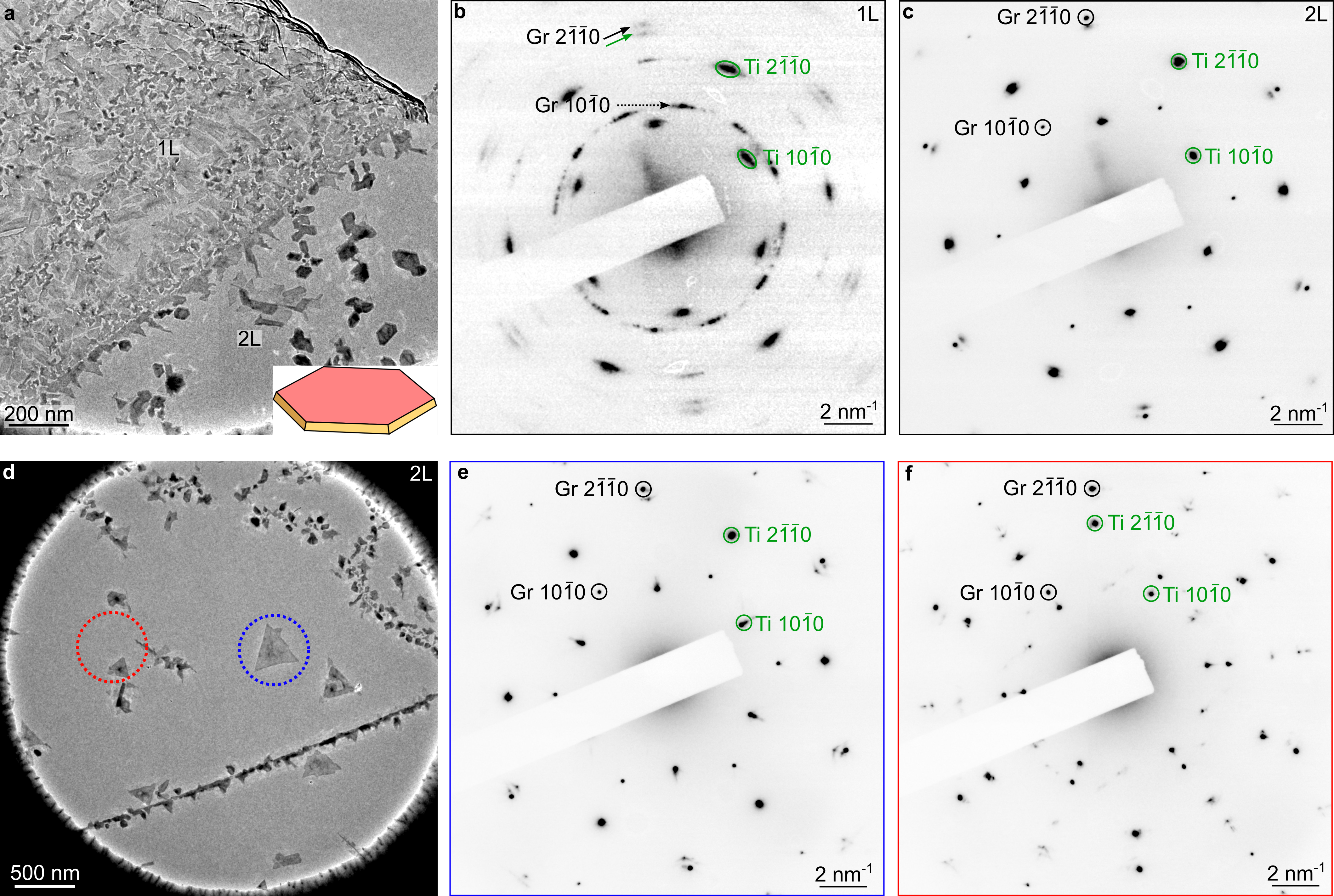}}
	\caption{\label{fig:400C}
		\textbf{Layer dependent Ti morphology after depositing Ti at 400 $\degree$C.} 
        \textbf{(a)} TEM image of a region with 1L and 2L Gr after Ti deposition at 400\,$\degree$C. The inset shows the calculated equilibrium Winterbottom shape of Ti which is bounded by the top (0001) surface (red) and the $\{11\bar{2}1\}$ side facets (gold).
        \textbf{(b, c)} Diffraction patterns from the (b) 1L  and (c) 2L region. Gr and Ti reflections are indexed with black and green font, respectively. The green arrow points to a second order Ti $\{10\bar{1}0\}$ spot.
        \textbf{(d)} TEM image of a different region on the same sample with 2L Gr. The linear feature is likely a fold in the Gr (see also Fig.~S6, caption).
        \textbf{(e, f)} Diffraction patterns from the regions indicated with the blue and red circle in (d).
        }
\end{figure*}

There is a marked difference in the Ti morphology between 1L and 2L or 3L Gr after a heated Ti deposition. The difference is more evident than that seen for room temperature depositions. On 2L and 3L Gr, heated deposition produces faceted islands that approach the Winterbottom equilibrium shape of hexagonally close-packed Ti (inset, Fig.~3(a); see \textit{Methods} for calculation details) \cite{winterbottom1967equilibrium}. It is typical for metallic islands to exhibit dendritic growth at lower temperatures and compact, faceted growth at higher temperatures, due to increased adatom diffusion along island edges as well as diminished kinetic limitations \cite{zhang1997atomistic}.  Our islands in fact exhibit a variety of morphologies, similar to the coexistence between different morphologies that is known for Pt and Al islands deposited at intermediate temperatures between 0.2-0.3~$T_{\mathrm{M}}$ (see Fig.~3.15 in \cite{michely2012islands}). We expect the morphology to converge towards the Winterbottom shape at temperatures above 0.3~$T_{\mathrm{M}}$: literature observations of Nb, Au, and Ti on few-layer Gr \cite{reidy2022mechanisms} also approached the Winterbottom morphology as the temperature was increased.

We also find that $N$ decreases with increasing temperature. We estimate $N_{\mathrm{2L,\,300\,\degree C}}$=100~µm$^{-2}$ and $N_{\mathrm{3L,\,300\,\degree C}}$=160~µm$^{-2}$ for 2L and 3L Gr, respectively, at a deposition temperature of 300\,$\degree$C. At a deposition temperature of 400\,$\degree$C, we measure $N_{\mathrm{2L,\,400\,\degree C}}$=12~µm$^{-2}$ on 2L Gr. This data was extracted from Fig.~S6. Such a decrease in $N$ is expected in a thermally activated diffusion process and suggests that defects do not dominate the diffusion properties of Ti on 2L or 3L~Gr. We did not extract values for $N$ for 1L Gr at 300\,$\degree$C due to the highly irregular shape of the Ti islands (see Fig.~S4-S6) and the fact that at 400\,$\degree$C, 1L Gr displays a nearly continuous Ti film from which nucleation sites cannot be discerned.

The diffraction pattern for 2L Gr (Fig.~3(b)) shows that the faceted Ti islands are epitaxially aligned with the Gr, although we find that for Ti deposited at 400\,$\degree$C, around 20\% of the islands are rotated 30$\degree$ with respect to the expected epitaxial alignment. Figure~S7 shows several images from the sample shown in Fig.~3 used to obtain statistics of the island rotations. Figure~3(d) highlights one overview TEM image from a 2L Gr region of this sample. The blue and red dotted circles indicate the location of the selected area aperture used to acquire the diffraction patterns shown in Fig.~3(e) and (f), respectively. The red dotted circle encloses a Ti island with a 30$\degree$ rotation. We believe that thermodynamic barriers to forming these less favorable configurations may be overcome by thermal energy at higher temperature. Since large scale defects are not observed in the rotated islands, we expect the rotated interface structure may be present at nucleation, and be preserved as the islands grow larger.

On 1L Gr we observe a different behavior compared to 2L and 3L Gr. In this case, heated deposition produces irregularly shaped islands. This can be understood if reactions with Gr hinder the free movement of the incoming Ti atoms. We have in several instances observed visible damage to the 1L Gr after heated Ti deposition, such as holes or rolled over Gr (see top right corner of the image in Fig.~3(a) and Fig.~S8 for higher-magnification images), indicating damage to 1L Gr during heated deposition. This is also consistent with Raman spectroscopy data, presented below, showing more defect formation and/or stronger Ti-Gr interaction on 1L Gr. Thus, these results indicate that the suspended 1L Gr becomes damaged when depositing Ti on Gr while heated to 300-400\,$\degree$C.

\begin{figure*}[t]
	\scalebox{\figurescale}{\includegraphics[width=0.66\linewidth]{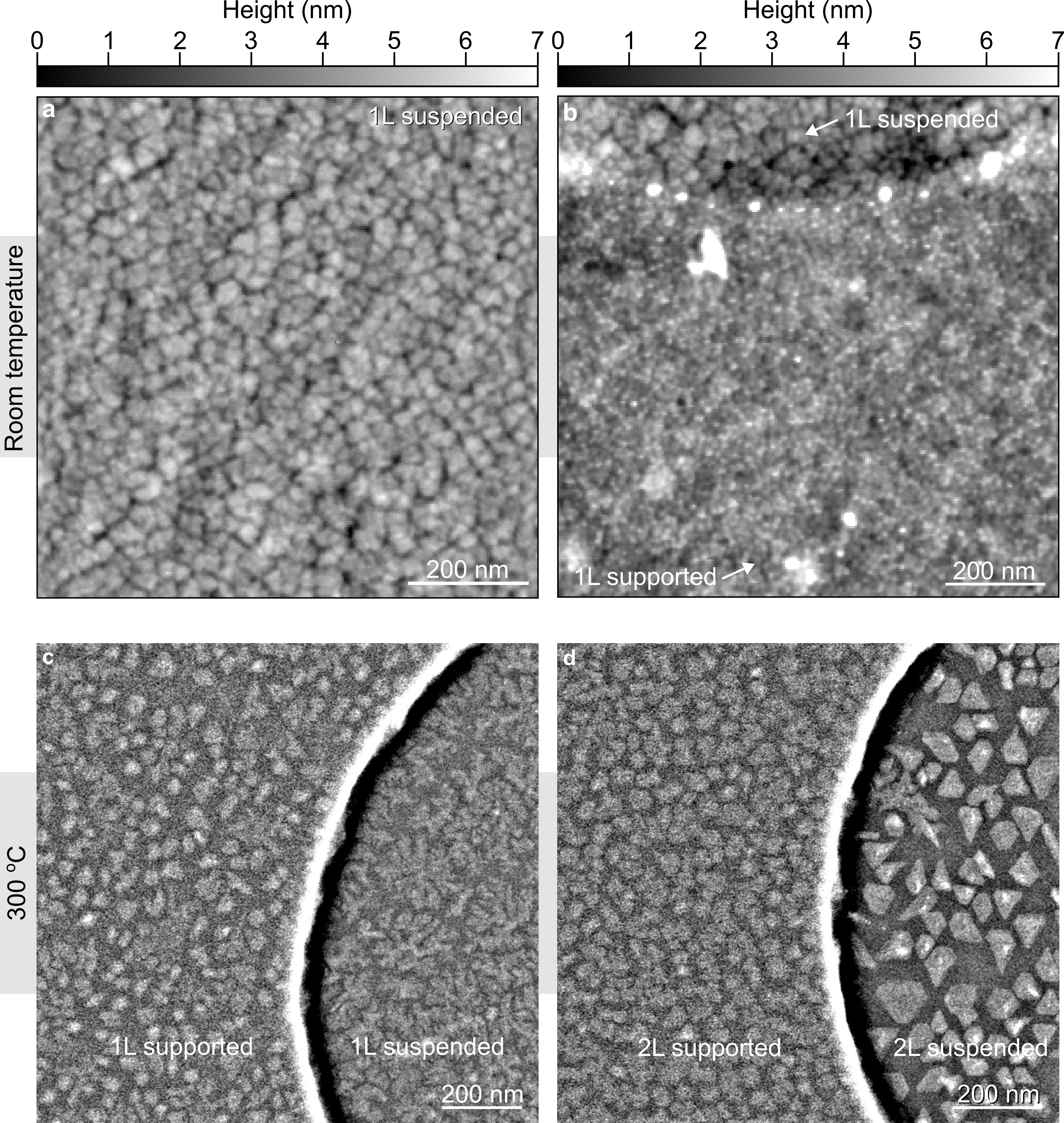}}
	\caption{\label{fig:timorph}
		\textbf{The effect of substrate on Ti morphology.} 
        \textbf{(a, b)} Ti depositions at room temperature. AFM scans of 1.1 nm Ti deposited onto regions with (a) suspended, (b) an area containing both suspended and SiN-supported 1L Gr as indicated with a white arrow. 
        \textbf{(c, d)} Ti depositions at 300\,$\degree$C. DF-STEM images of (c) 1L Gr and (d) 2L Gr.}
\end{figure*}

The diffraction pattern for 1L Gr (Fig.~3(b)) also indicates defects in the Gr. There are additional Gr reflections at the positions corresponding to the Gr $\{10\bar{1}0\}$ d-spacing, which indicate the formation of polycrystalline Gr. In reaching this conclusion, we also considered the possibility of a chemical reaction which forms a material with atomic spacing similar to that of the Gr $\{10\bar{1}0\}$  reflections. The most likely candidates would be TiC or TiO$_2$, discussed in Table S2 and its caption. However, comparing our diffraction patterns with expected diffraction reflections from these compounds does not yield a close fit. Other possibilities could include ternary compounds of C, Ti, and O, or the formation of a non-equilibrium compound. However, DF-TEM imaging of 1L Gr shows that the reflections from the diffuse Gr ring do not correlate well with the locations of the Ti islands (Fig.~S9). Overall, therefore, we believe the additional reflections at the Gr $\{10\bar{1}0\}$  are most likely due to damage to the Gr and the formation of polycrystalline Gr. An additional observation is that the Ti $\{10\bar{1}0\}$  and $\{2\bar{1}\bar{1}0\}$ spots are more diffuse on suspended 1L Gr compared to 2L Gr, indicating a larger variation in crystal orientations. There is also a second set of faint spots by the Gr $\{2\bar{1}\bar{1}0\}$ spots at a slightly smaller radius (marked with a green arrow, Fig.~3(b)). These match precisely with the second order Ti $\{10\bar{1}0\}$ reflections.

In summary, we find that the result of high temperature deposition depends strongly on the Gr thickness. For 2L and 3L Gr, depositing Ti at 300-400\,$\degree$C results in the formation of faceted and compact islands with morphologies close to the Winterbottom shape, as expected when thermodynamic factors determine island shape. Furthermore, $N$ decreases with temperature as expected for a thermally driven diffusion process, indicating that it is not governed by a few low-energy defect sites on the Gr. In contrast, on 1L Gr, depositions at 300\,$\degree$C or 400\,$\degree$C lead to irregularly shaped islands which can be explained by strong interactions with the Gr and defect formation. 

\subsection{Substrate dependence of Ti morphology and nucleation characteristics}

In most technologically relevant applications, the Gr is supported on a substrate rather than used as a freestanding film. We showed previously that Gr roughness affects Ti nucleation, and it is known that substrates such as SiO$_2$ or SiN increase the surface roughness of Gr \cite{ishigami2007atomic}. We therefore now discuss Ti morphology on SiN-supported Gr, comparing with freestanding Gr to investigate substrate effects. Figure 4(a,\,b) compares atomic force microscopy (AFM) images of SiN-supported and suspended 1L Gr with Ti deposited at room temperature. SiN-supported 1L Gr shows a larger nucleation density and smaller island size compared to suspended 1L Gr, to such an extent that identifying individual islands becomes challenging.

We again interpret these results based on differences in surface roughness. In literature for Gr supported on SiO$_2$ and SiN, the root mean square roughness has been measured to be 0.19-0.35~nm, attributed to roughness imposed from these uneven surfaces \cite{ishigami2007atomic, cullen2010high, yamamoto2012charge, geringer2009intrinsic}. This is larger than the intrinsic roughness experienced by suspended 1L Gr, which in literature has been measured to be about 0.1~nm \cite{thomsen2017suppression, gao2014thermomechanics, ramirez2016anharmonic}. Therefore, we would expect lower Ti diffusivity on Gr supported by a substrate, which in our case results in a very large nucleation density. 

We then compare the Ti island morphology between suspended and SiN-supported Gr after deposition at 300\,$\degree$C. Figure 4(c,\,d) shows dark field-STEM images of regions containing both suspended and supported 1L and 2L Gr, respectively. Suspended 1L Gr displays irregularly shaped islands as expected. Interestingly, the SiN-supported 1L Gr displays compact islands which suggests less defect formation in the Gr and a weaker Ti-Gr interaction in these regions. 

On 2L Gr, the Ti nucleation density is larger on SiN-supported Gr compared to suspended Gr. This is again indicative of smaller surface roughness on the suspended region, and similar to results obtained in a different study with Au deposition on Gr \cite{thomsen2022suspended}. We rule out temperature differences between suspended and supported 1L Gr as the reason for the observed differences in Ti island morphology. This is because the whole TEM sample carrier is heated resistively during heated depositions, and we would expect a uniform temperature in the small area where the Gr is located. 

Hence, we conclude that the presence of a substrate increases surface roughness, resulting in a larger nucleation density of islands. This is seen for both 1L Gr at room temperature and for 2L Gr at elevated temperatures. Depositions on SiN-supported 1L Gr at elevated temperature result in the formation of compact Ti islands, indicating that the presence of a substrate leads to less defect formation compared to depositions on 1L suspended Gr at elevated temperature where irregularly shaped islands form.  

\subsection{Defect formation in Gr during Ti deposition}

So far, we have discussed nucleation density and Ti island morphology, which are macroscopic phenomena. Now we discuss a key nanoscale structural feature, defects in Gr, which are typically studied by Raman spectroscopy. 

\begin{figure*}[t]
	\scalebox{\figurescale}{\includegraphics[width=0.67\linewidth]{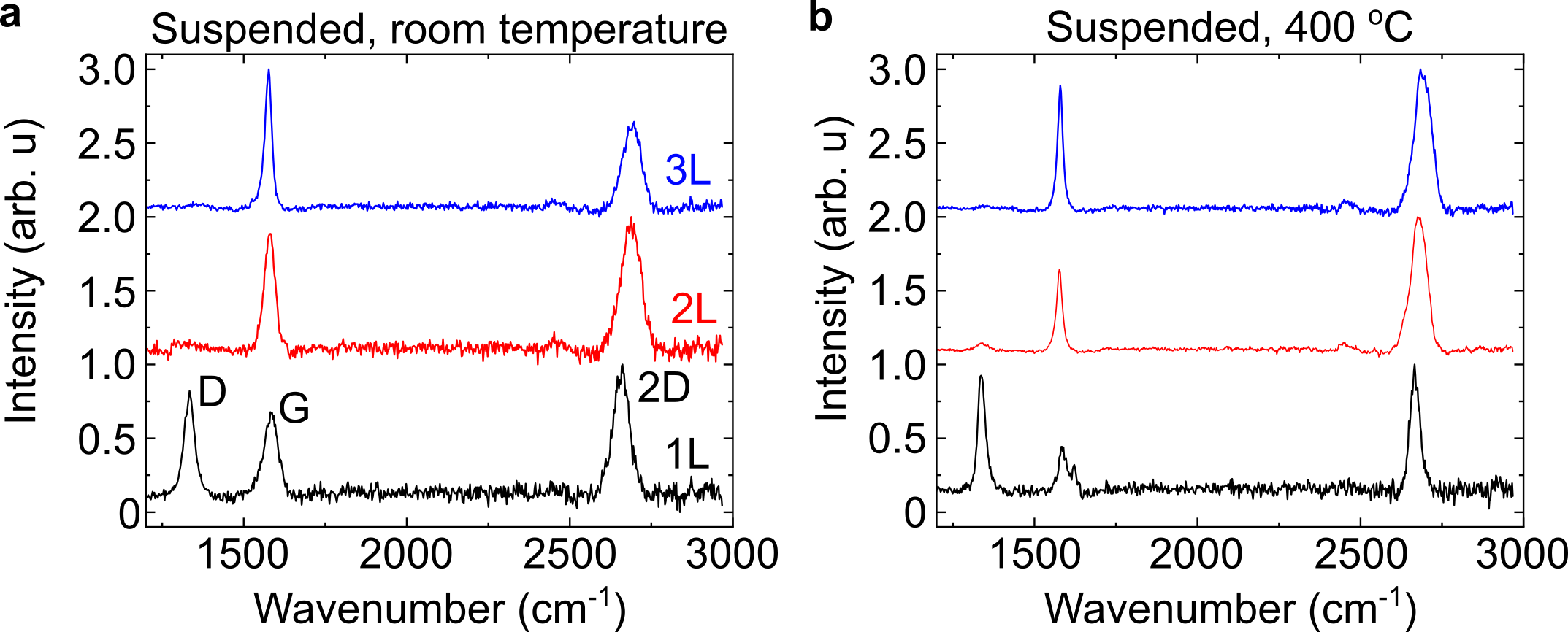}}
	\caption{\label{fig:raman}
		\textbf{Raman spectroscopy.} 
        \textbf{(a)} After Ti deposition on suspended Gr without intentional heating. 
        \textbf{(b)} After Ti deposition on suspended Gr heated to 400\,$\degree$C.}
\end{figure*}

To evaluate defect formation in the Gr during Ti evaporation we obtained Raman spectroscopy data from samples with 1L, 2L, and 3L suspended Gr after Ti deposition at room temperature and 400\,$\degree$C (Fig.~5). Raman spectra of Gr show characteristic G and 2D peaks at $\sim$1580~cm$^{-1}$ and 2690~cm$^{-1}$, respectively. Literature indicates that the D-peak at $\sim$1350~cm$^{-1}$ appears only in the presence of certain defects. The D-peak is associated with vacancy-type and sp$^{3}$-type defects, which occur when Gr forms bonds to other materials or compounds, such as TiC bonds \cite{eckmann2012probing}. Charged impurities, intercalants and strain do not produce a D-peak \cite{ferrari2013raman}. In Gr devices, the intensity of the D-peak, typically measured as the intensity ratio between the D- and the G-peak, $I_{\mathrm{D/G}}$, has been found to scale inversely with the charge carrier mobility \cite{chen2009defect, ni2010resonant}. This underscores the importance of strategies for avoiding defect formation during processing when Gr is present.

In our case, suspended 1L Gr with Ti deposited at room temperature exhibits a D-peak, while 2L and 3L Gr do not show a noticeable D-peak. This suggests that defect formation, such as vacancy-type and sp$^3$-type defects, is strongly reduced or even eliminated for 2L and 3L Gr. Similarly, Ti deposition at 400\,$\degree$C leads to a D-peak on 1L Gr, which is relatively larger than that from the room temperature deposition.

The defect concentration is quantified using the intensity ratio between the D- and the G-peak, $I_{\mathrm{D/G}}$ \cite{lucchese2010quantifying}. For 1L Gr at room temperature ($I_{\mathrm{D/G}}$=1.3), the defect density is 1.4$\cdot$10$^4$~µm$^{-2}$. With a Ti nucleation density of 1280~µm$^{-2}$ on 1L suspended Gr at room temperature this suggests that each Ti island contains $\sim$11 point defects in the underlying 1L Gr support. For 1L Gr at 400\,$\degree$C ($I_{\mathrm{D/G}}$=2.9), the defect density increases to $5.4\cdot10^4$~µm$^{-2}$. For 2L Gr at 400\,$\degree$C ($I_{\mathrm{D/G}}\sim$0.1), we estimate an upper bound of $<1.6\cdot10^3$~µm$^{-2}$ defects. This upper bound is based on measurements in Ref.\ \cite{lucchese2010quantifying},  which considers a $I_{\mathrm{D/G}}$-ratio down to 0.2, corresponding to this defect density. These results indicate a strong Gr thickness and temperature dependence on the defect formation in Gr after Ti deposition. 

If we compare these Raman spectroscopy data with the structural characterization from Fig.~1 and 3, we obtain a more complete picture of the defect structure of the samples. For room temperature depositions, 1L Gr exhibits many more defects in Raman spectroscopy than 2L or 3L Gr, although structural changes (nucleation density and island size) change more gradually with Gr layer number. In general, high-temperature depositions show more defects compared to room temperature depositions. For high temperature depositions, 1L Gr shows at least an order of magnitude higher defect density than 2L or 3L Gr. For the high temperature case, the structural changes are also dramatic: 1L Gr displays irregularly shaped Ti islands and visible holes, whereas 2L and 3L Gr show compact and faceted islands. 

Our previous analysis showed that increasing surface roughness results in a stronger Ti-Gr interaction, and 
we believe the differences in defect formation for 1L-3L Gr are also related to surface roughness. The reactivity of Gr has been shown to increase with increasing surface roughness \cite{liu2008graphene, boukhvalov2009enhancement, fan2011effect}. Since surface roughness of free-standing Gr decreases with increasing thickness \cite{meyer2007roughness}, our Raman and structural TEM characterization align well with these previous studies. These results also suggest that placing 1L Gr on a flat substrate like hexagonal boron nitride or another relevant 2D material could be an avenue towards reducing defect formation during Ti layer growth.

\section{Conclusions}
To conclude, we report a complicated dependence of Ti-Gr interface morphology and defect formation on Gr layer thickness, substrate, and temperature, and we discuss how these can be understood through the central role of Gr roughness. Our findings are summarized in Table 2. Ti island nulceation density, shape and degree of faceting depend on the temperature in a way expected from kinetic considerations, but these parameters also depend on Gr thickness. Since DFT simulations do not predict a strong dependence of the Gr-Ti interaction on Gr thickness, we conclude that an extrinsic factor, roughness, dominates the island growth process.

\begin{table*}[t]
\centering
\begin{tabular}{p{2.5cm} p{8.2cm} p{0.3cm} p{6cm} p{0cm}}\hline
\centering \textbf{Parameter} & \centering \textbf{Effect of Ti morphology} & & \centering \textbf{Effect on Gr damage} & \\ \hline 
\centering Gr layer number & Increasing Gr thickness leads to larger Ti diffusivity on the Gr surface, resulting in lower nucleation density and larger Ti islands (Fig.~1 \& 2). & &  Damage (as identified through Raman spectroscopy) decreases with Gr thickness (Fig.~5). & \\  
\centering Temperature & Increasing the temperature diminishes kinetic factors and increases thermodynamic factors in the growth of Ti islands, resulting in faceted islands with a shape predicted by the Winterbottom reconstruction (Fig.~3). & &  Increasing the temperature results in more damage to the Gr (Fig.~5). & \\
\centering Substrate &  A substrate like SiN reduces Ti diffusivity due to increased surface roughness resulting in a continuous film with small grain sizes when depositing Ti at room temperature. At elevated temperature we also observe an increased nucleation density on supported Gr compared to free-standing Gr (Fig.~4). & & Faceted islands on SiN-supported 1L Gr (as opposed to the irregular islands observed on suspended 1L Gr) after Ti deposition at elevated temperatures suggest less defect formation when a substrate is present (Fig.~4).  & \\ \hline
\end{tabular}
\caption{Summary of experimental results for the dependence of Ti deposition on growth and substrate parameters.}
\label{tab:2}
\end{table*}

Results for 1L Gr stand out from the results of 2L or 3L Gr. 1L Gr displays at least an order magnitude larger defect density and completely different Ti island morphologies as compared to thicker Gr flakes. This supports the idea of these differences arising due to structural and mechanical differences between 1L and thicker Gr flakes, where 1L Gr is more flexible, exhibits larger surface roughness and, therefore, displays lower Ti diffusivity and is more easily damaged.

These results highlight the extreme sensitivity of processing conditions on Ti deposition on Gr and emphasize the importance of considering factors such as the Gr thickness, the substrate on which Gr is placed, particularly its roughness, for applications where the microscopic nature of the metal-Gr interface is critical to performance. This is in addition to the usual factors considered in metal nucleation and growth, such as deposition rate and substrate temperature. 

The summary in Table 2 offers guidelines for creating Ti-Gr interfaces with ideal properties. For large single crystals with minimized defect formation, it would be interesting to investigate Ti growth on Gr supported by hexagonal boron nitride or another thicker 2D material, as this should suppress damage formation, allowing the growth of faceted islands, while also providing a flat substrate minimizing nucleation density and creating large islands. It may be further possible to grow large, faceted islands by increasing the temperature. 
Another option is a two-step process with a low-temperature initial deposition to limit interfacial defect formation, followed by a higher temperature deposition to increase Ti grain size.
Additionally, it would be interesting to extend this work to other 2D materials like transition metal dichalcogenides (TMDs), where the Ti-TMD interface is heavily utilized in nanoelectronics research and control of the reactions at the interface are desired \cite{schauble2020uncovering, mcdonnell2016mos2}. 
Future studies should also examine the properties of Ti as an adhesion layer to other metals, and how the properties of other metals are affected by the Ti layer. It may be possible to grow a metallic overlayer which is epitaxial with respect to the epitaxial Ti film, and therefore create epitaxial multilayers on Gr. Furthermore, the faceted Ti islands obtained by evaporating Ti at elevated temperatures may be useful for optoelectronic applications and may serve as seeds for other metals on Gr to create complicated multilayer island structures.

\section{Methods/Experimental}

\footnotesize{

\subsection{Sample Fabrication}
Gr was obtained by exfoliating natural graphite (NaturGrafit GmbH, Germany) onto oxygen plasma treated substrates of 90 nm thick SiO$_2$ on Si using 3M Magic Scotch tape. We identified suitable flakes by their contrast using optical microscopy as shown in Fig.~S1. We then transferred crystals to either homemade TEM compatible sample carriers or "location-tagged TEM grids" purchased from Norcada, Inc., Canada. The homemade grids consist of 300~nm thick SiN membrane windows with multiple $\sim$3~µm diameter holes etched in the window. The Norcada TEM grids have a 200~nm thick SiN membrane with several 2~µm holes. The Gr is transferred using wedging transfer, based on cellulose acetate butyrate as polymer handle for the transfer, which we have previously shown produces ultra-clean suspended Gr after thermal annealing in vacuum \cite{thomsen2022suspended, thomsen2017suppression}. After the transfer we bake the TEM sample carriers to 140~$\degree$C to improve adhesion. Then we dissolve the polymer in acetone and rinse the TEM sample carriers in isopropanol. When drying the samples we use critical point drying. This is typically used for fragile samples where surface tension could otherwise cause damage, including free-standing Gr samples \cite{thomsen2022suspended, reidy2022mechanisms, thomsen2017suppression, meyer2007structure, bolotin2008ultrahigh}. Damage to Gr during a regular drying process, where a solvent like isopropanol evaporates from the surface of the TEM grid, typically results in easily observed (through optical microscopy) macroscopic effects such as torn or rolled Gr, especially in 1L Gr.

\subsection{Ti deposition}
For the suspended Gr, to remove polymeric residue, we anneal the sample overnight at 500\,$\degree$C prior to evaporation. The annealing takes place in an ultrahigh vacuum chamber that also contains the Ti evaporator. The pressure during the evaporation was typically between $1\times10^{-8}$--$2\times10^{-8}$ Torr. We deposit Ti using an electron-beam evaporator at a typical rate of around 0.3~Å/min and a thickness of approximately 1~nm. Fig.~S10 shows Raman spectroscopy data obtained after Gr transfer to the TEM grids ("as prepared"), after the annealing step in our UHV system ("after UHV anneal"), and after Ti deposition at 400\,$\degree$C. This data shows that Gr transfer and UHV annealing does not lead to any detectable defect formation in the Gr, and that a D-peak in 1L Gr only appears after Ti deposition.

\subsection{Electron microscopy}
TEM images and diffraction patterns were obtained from a JEOL 2010 TEM operated at 200~kV. A selected area aperture with a diameter of 520~nm was used for acquiring diffraction patterns. DF-STEM and TEM images were obtained from a Hitachi HF5000 environmental TEM operated at 200~kV. 

\subsection{Data analysis}
Ti island nucleation density was estimated by delineating Ti islands by hand and measuring their areas using imageJ \cite{thomsen2022suspended}. Fig.~S3 and S6 show these measurements. Extracting the nucleation density for our depositions at elevated temperatures is challenging since at least 50\% of islands have coalesced.

\subsection{Raman spectroscopy}
Raman spectroscopy was performed using a Renishaw inVia confocal Raman microscope with a 532~nm excitation laser. We acquired maps with 0.5~µm spacing between spectra. We then sum 10 spectra to obtain the data shown in Fig.~5. The laser power was set to 2.5 mW. This is well below the power required to visibly damage the Gr.

\subsection{Atomic force microscopy}
AFM was performed on a Cypher VRS AFM operated in tapping mode. 

\subsection{Winterbottom reconstruction calculations}
We calculated the equilibrium Winterbottom shape of Ti nanoislands on Gr using a Mathematica interface \cite{reidy2022mechanisms, winterbottom1967equilibrium, zucker2012new}. We obtained the relative surface energies of the crystal facets for HCP Ti from high-throughput density functional theory \cite{mavrl, kirklin2015open, tran2019anisotropic, zheng2020grain}. The Winterbottom reconstruction shown in 3(b) was obtained using the following interface energies (all given in J/m$^2$): $\{0001\}$: 2.15; $\{11\bar{2}1\}$: 1.93; $\{2\bar{1}\bar{1}2\}$: 2.01; $\{11\bar{2}0\}$: 2.04. A more quantitative value for the interface energy could not be obtained, since the Ti islands do not form their full equilibrium geometries, however the overall shape of the islands tends closer towards equilibrium upon heated deposition.

\subsection{Density Functional Theory Calculations}
The DFT calculations are carried out using VASP 5.2 \cite{kresse1996efficient} employing the Perdew, Burke, and Ernzerhof (PBE) Generalized Gradient Approximation (GGA) \cite{perdew1996generalized} for the exchange-correlation function and PAW pseudopotentials \cite{blochl1994projector, kresse1999ultrasoft}. Van der Waals interactions are accounted for using the Tkatchenko-Scheffler correction \cite{tkatchenko2009accurate, al2012assessment}. The electronic wavefunctions are expanded with planewaves with energy less than 400~eV. The electronic self-consistent loop is terminated when energy changes are less than $1\cdot10^{-6}$ eV and the ionic relaxations are considered converged when the magnitude of the largest force on any atom is less than 0.01~eV/Å. The Gr slab is modeled using a 3$\times$3 surface supercell created from graphite with hexagonal lattice constants 2.458 and 6.592~Å. For the Ti layers, we used a slab comprising of 15 layers, where the top 6 layers (those furthest from the Gr) were fixed at their bulk positions to minimize finite-size effects. For simplicity, Fig.~1(f) shows only the first 4 Ti layers. To mitigate the fictitious interactions between the images in the non-periodic direction perpendicular to the Gr slab, we used a supercell approach with at least 12~Å of vacuum. The Brillouin zone is sampled using the 5$\times5\times$1 Monkhorst-Pack grid. All of the calculations are spin-polarized. To determine the interface structure, we used the Zur and McGill \cite{zur1984lattice} lattice matching scheme in conjunction with lattice shifs \cite{guo2017structural, guo20172d}. The lattice matching scheme accounts for possible rotations of one lattice with respect to the other, or expansion of both lattices, in addition to the lattice reduction scheme, to minimize the strain and lattice mismatch between the two lattices. The precision of the match gives an upper bound on the strain required to form chemical bonds across the interface. The lattice shifts ensure that the atom registry at the interface is optimum \cite{guo2017structural}. 

\section{Supporting Information}
The Supporting Information is available free of charge at [insert link here]. 

Optical microscopy sample overview images (Fig.~S1), high-magnification TEM images of 1L Gr with Ti deposited at room temperature (Fig.~S2), nucleation density measurement data at room temperature (Fig.~S3), additional TEM data for Ti depositions at elevated temperature including nucleation density measurements and high-magnification images (Fig.~S4-S8), DF-TEM images of 1L Gr with Ti deposited at 400\,$\degree$C, Raman spectroscopy data obtained during sample fabrication (Fig.~10), additional DFT results (Table~S1), scattering data from x-ray scattering for Gr, Ti, TiC, and TiO$_2$ (Table~S2).

\section{Acknowledgements}
This work is primarily supported through the U.S. Department of Energy, Office of Science, Basic Energy Sciences (BES), Materials Sciences and Engineering Division under FWP ERKCK47 ‘Understanding and Controlling Entangled and Correlated Quantum States in Confined Solid-state Systems Created via Atomic Scale Manipulation’, as well as partially supported by the Quantum Science Center (QSC), a National Quantum Information Research Center of the U.S. Department of Energy (DOE) on probing quantum matter. W.A.S.\ is partially supported by the US National Science Foundation (award no. CBET-2130804). Computational support was provided by the University of Pittsburgh Center for Research Computing through the resources provided on the H2P cluster, which is supported by NSF (award no.\ OAC-2117681). J.D.T.\ acknowledges support from Independent Research Fund Denmark through grant no.\ 9035-00006B. K.R.\ acknowledges funding from the Hugh Hampton Young Memorial Fund at MIT. J.J.P.\ acknowledges funding from the Natural Sciences and Engineering Research Council of Canada and MIT School of Engineering. The authors would like to acknowledge Michael Tarkanian for help in manufacturing TEM sample holders and Prof. Jeehwan Kim for equipment access. This work made use of the Materials Research Laboratory Shared Experimental Facilities at MIT, supported in part by the MRSEC Program of the National Science Foundation under award no.\ DMR1419807. This work was carried out in part using MIT.nano's facilities. Figure 1(d) and 2(c) were created using Vesta \cite{momma2011vesta}.

\bibliography{references}
\bibliographystyle{naturemag}%

\end{document}